\documentclass[showpacs,twocolumn,preprintnumbers,amsmath,amssymb]{revtex4}
\usepackage{amsmath,amsfonts,latexsym,amssymb,graphicx,graphics,epsfig,subfigure,color,makeidx}
\usepackage{xcolor,diagbox}
\usepackage{multirow}
\usepackage[colorlinks,linkcolor=blue,anchorcolor=blue,citecolor=green,urlcolor=blue]{hyperref}
\usepackage{mathrsfs}
\usepackage{amsmath}
\newcommand {\nn}{\nonumber}

\begin{document}
\title{Quasibound and quasinormal modes of a thick brane in Rastall gravity}

\author{Qin Tan$^{a}$\footnote{tanqin@hunnu.edu.cn}}
\author{Yi Zhong$^{b}$\footnote{zhongy@hnu.edu.cn}}
\author{Wen-Di Guo$^{c}$$^{d}$\footnote{guowd@lzu.edu.cn, corresponding author}}

\affiliation{
$^{a}$Department of Physics, Key Laboratory of Low Dimensional Quantum Structures and Quantum Control of Ministry of Education, Synergetic Innovation Center for Quantum Effects and Applications, Hunan Normal University, Changsha, 410081, Hunan, China\\
$^{b}$School of Physics and Electronics Science, Hunan University, Changsha 410082, China\\
$^{c}$Lanzhou Center for Theoretical Physics, Key Laboratory of Theoretical Physics of Gansu Province, Lanzhou University, Lanzhou 730000, China\\
$^{d}$Institute of Theoretical Physics and Research Center of Gravitation, School of Physical Science and Technology, Lanzhou University, Lanzhou 730000, China}

\begin{abstract}
In this work, we study the gravitational quasinormal modes of the thick brane in Rastall gravity. Using the asymptotic iteration and direct integration methods, we solve the quasinormal frequencies of the Rastall thick brane. We also obtained the waveforms of these quasinormal modes through numerical evolution. The results indicate that although the Rastall thick brane lacks a bound zero mode, when the Rastall parameter $\lambda\gtrsim0$, a long-lived quasinormal mode appears. This long-lived quasinormal mode may restore the four-dimensional effective Newtonian potential on the brane on a large scale. This may provide a new perspective for the localization of gravity on thick branes, that a thick brane does not necessarily require the gravity to be localized, perhaps quasi-localized is sufficient.

\end{abstract}

\maketitle

\section{Introduction}
\label{Introduction}
The concept of extra dimensions has been around for over a century~\cite{kaluza:1921un,Klein:1926tv}, and in the past two decades, the braneworld theory has attracted considerable attention. The initial purpose of braneworld theory is to address the hierarchy problem in particle physics by introducing additional spatial dimensions. Some prominent models in this theory include the large extra dimension model proposed by Arkani-Hamed $et~al.$~\cite{Arkani-Hamed:1998jmv,Antoniadis:1998ig}, and the warped extra dimension model proposed by Randall and Sundrum (RS-I)~\cite{Randall:1999ee}. Building upon the RS-I model, Randall further proposed the RS-II braneworld model~\cite{Randall:1999vf}, which features infinite large extra dimensions while still recovering the Newtonian potential on the brane. The emergence of the RS-II model has challenged people's inherent understanding of the scale of extra dimensions, and its theory has been closely integrated with black hole physics, cosmology, particle physics, and other fields, yielding significant achievements and advancements~\cite{Shiromizu:1999wj,Tanaka:2002rb,Gregory:2008rf,Jaman:2018ucm,Adhikari:2020xcg,Bhattacharya:2021jrn,Geng:2020fxl,Geng:2021iyq}. Braneworld models like RS model, due to the infinitesimal thickness of the brane, are referred to as thin brane models. Considering that the brane has thickness along the extra dimension, DeWolfe $et~al.$ proposed the thick brane models~\cite{DeWolfe:1999cp,Gremm:1999pj,Csaki:2000fc}. The thick brane model is a smooth extension of the RS-II model, with a rich internal structure. In previous studies, the solutions of thick brane in various gravity theories and the localization of matter fields on thick branes have been investigated~\cite{Afonso:2007gc,Dzhunushaliev:2010fqo,Dzhunushaliev:2011mm,Geng:2015kvs,Melfo2006,Almeida2009,Zhao2010,Chumbes2011,Liu2011,Bazeia:2013uva,Xie2017,Gu2017,ZhongYuan2017,ZhongYuan2017b,Zhou2018,Hendi:2020qkk,Xie:2021ayr,Moreira:2021uod,Xu:2022ori,Silva:2022pfd,Xu:2022gth}. More information about the development of the thick brane can be found in these reviews~\cite{Liu:2017gcn,Ahluwalia:2022ttu,Dzhunushaliev:2009va}.

Recently, reference~\cite{Zhong:2022wlw} investigated the thick brane model in Rastall gravity. Rastall gravity is a modified gravity proposed by P. Rastall where the energy-momentum tensor is not conserved~\cite{Rastall:1972swe}. While conservation of energy-momentum tensor is a fundamental assumption in general relativity and most modified gravity theories, there is no clear experimental evidence supporting this in curved spacetime. Particle creation processes in cosmology also violate energy-momentum conservation~\cite{Parker:1969au,Gibbons:1977mu,Ford:1986sy}. Therefore, Rastall gravity relaxes the requirement of energy-momentum conservation and has been extensively studied in cosmology and compact stars~\cite{BezerradeMello:2014okn,Heydarzade:2017wxu,Darabi:2017tay,Xu:2017vse,Darabi:2017coc,Das:2018dzp,Tang:2019dsk,Li:2019jkv,Khyllep:2019odd,Ghosh:2021byh,Haghani:2022lsk,Shahidi:2021lxt}. The connection between Rastall gravity and $f(R,T)$ gravity has also been explored~\cite{Shabani:2020wja}. In the Rastall thick brane, the flat brane case does not possess a zero mode of gravitational perturbation~\cite{Zhong:2022wlw}. In most braneworld models, to recover the four-dimensional effective Newtonian potential, the gravitational zero mode must be bound to the brane. Can the four-dimensional effective Newtonian potential on the Rastall brane still be restored?

In some thin brane models, like the Dvali-Gabadadze-Porrati (DGP) model~\cite{Dvali:2000hr} and Gregory–Rubakov–Sibiryakov (GRS) model~\cite{Gregory:2000jc} cannot localize the gravitational zero mode on the brane. Instead, these models restore the Newtonian potential on the brane within a specific screening radius through a quasi-localized gravitational resonant mode. Does a similar situation exist for a thick brane model? Since there are no zero mode in the static flat thick brane solution of Rastall gravity, the question arises: where do the zero modes have gone? In our previous study~\cite{Tan:2022vfe}, it has been found that besides zero modes, thick branes also possess a series of discrete quasinormal modes (QNMs), and these QNMs have a profound connection with the resonant states of the thick brane~\cite{Tan:2023cra}. Quasinormal modes are characteristic modes of dissipative systems containing crucial information about the system. They have been extensively studied in areas like black hole physics, leaky resonant cavities, and brane worlds~\cite{Berti:2009kk,Kokkotas:1999bd,Nollert:1999ji,Konoplya:2011qq,Cardoso:2016rao,Jusufi:2020odz,Cheung:2021bol,Kristensen:2015qq,Seahra:2005wk,Seahra:2005iq}. Starting from the QNMs of the Rastall thick brane, it might be possible to reveal some characteristics of the Rastall thick brane and answer the above questions. Therefore, in this paper, we will investigate the properties of QNMs of the Rastall thick brane from both the frequency domain and time domain.

This paper is organized as follows. In Sec.~\ref{background and perturbation equation}, we review the thick brane solution in Rastall gravity and its linear tensor perturbation. In Sec.~\ref{quasibound and quasinormal modes of rastall brane},  we investigate the QNMs of the Rastall thick brane using asymptotic iteration method (AIM), direct integration method, and numerical evolution method in both frequency domain and time domain. We study the influences of the parameter $\lambda$ on the quasinormal spectrum and the instability of the thick brane. Finally, the conclusions and the discussions are given in Sec.~\ref{Conclusion}.

\section{background and perturbation equation}
\label{background and perturbation equation}
In this section, we will review the thick brane and its gravitational perturbation in Rastall gravity. In this theory, where the covariant divergence of the energy-momentum tensor is modified as follows~\cite{Rastall:1972swe}
 \begin{eqnarray}
	\nabla^{M}T_{MN}=\lambda \nabla_{N}R.	\label{covdT}
\end{eqnarray}
This modification leads to a revised field equation in five-dimensional spacetime, given by:
\begin{eqnarray}
	\label{Rastal eq1}
	R_{MN}-(\frac{1}{2}-\lambda)R g_{MN}=T_{MN}.
\end{eqnarray}
Here, the parameter $\lambda$ is called Rastall parameter, and current observations indicate that $|\lambda|\ll1$~\cite{Akarsu:2020yqa}. In this paper, we set the gravitational constant $\kappa=1$ for convenience. Clearly, when the parameter $\lambda\rightarrow0$, it goes back to general relativity. Hereafter, the capital Latin letters such as $M,N,\dots$ represent the five-dimensional indices with $M,N = 0,1,2,3,4$, while Greek letters like $\mu,\nu\dots$ represent the four-dimensional indices with $\mu,\nu\dots=0,1,2,3$, and Latin letters $i,j\dots=1,2,3$ are used for the three-dimensional spatial indices on the brane. We further consider that the thick brane is generated by a scalar field $\varphi$. Its energy-momentum tensor is given by
\begin{eqnarray}
	\label{energymomentum1}
T_{MN}=\partial_{M}\partial_{N}\varphi-\frac{1}{2}g_{MN}\left(\partial^{A}\varphi\partial_{A}\varphi-V(\varphi)\right).
\end{eqnarray}
Thus, Eq.~\eqref{Rastal eq1} is transformed into~\cite{Zhong:2022wlw}
\begin{eqnarray}
	\label{Rastal eq}
	R_{MN}-(\frac{1}{2}-\lambda)R g_{MN}&=&-\frac{1}{2}g_{MN}\left(\partial^{A}\varphi\partial_{A}\varphi-V(\varphi)\right)\nn\\
	&&+\partial_{M}\partial_{N}\varphi.
\end{eqnarray}
The metric is chosen as static flat brane metric:
\begin{equation}
	ds^2=e^{2A(z)}(\eta_{\mu\nu}dx^\mu dx^\nu+dz^2),
	\label{metric}
\end{equation}
where $e^{A(z)}$ denotes the warp factor and $\eta_{\mu\nu}=\text{diag}(-1,1,1,1)$ represents the four-dimensional Minkowski metric. Substituting the metric~\eqref{metric} into Eqs.~\eqref{covdT} and \eqref{Rastal eq}, we obtain the specific  dynamical equations as
    \begin{eqnarray}
	\label{eom1}
	&&3A'^2-3A''-\varphi'^2=0, \\
	\label{eom2}
	&&2\text{e}^{2A}V(\varphi)+ 3(3+8\lambda)A'^2+(3+16\lambda)A''=0,\\
	\label{eom3}
	&&-24\lambda A'^3+8 \lambda  A''' +4A'\left(2 \lambda A''+\varphi'^2\right)-\text{e}^{2A}\varphi' V'(\varphi )=0,\nonumber\\
\end{eqnarray}
where the prime denotes the derivative with respect to $z$. The thick brane solution was studied in Ref.~\cite{Zhong:2022wlw}:
    \begin{eqnarray}
	\label{solutionwarpfactor1}
A(z)&=&-\frac{1}{2}\ln(k^2 z^2 +1), \\
	\label{solutionphi1}
\varphi(z)&=&\sqrt{3}\text{arctan}(k z),\\
	\label{solutionV1}
	V(\varphi)&=&\frac{1}{4} k^2 \left[(56 \lambda +15) \cos \left(\frac{2 \varphi }{\sqrt{3}}\right)-3 (8 \lambda +3)\right],\nonumber\\
\end{eqnarray}
where the parameter $k$ has mass dimension one. Next, we consider the tensor perturbation of the metric of the Rastall brane. The perturbed metric is
\begin{eqnarray}
	g_{MN}=\left(
	\begin{array}{cc}
		e^{2A(z)}(\eta_{\mu\nu}+h_{\mu\nu}) & 0\\
		0 & e^{2A(z)}\\
	\end{array}
	\right)\label{perturbed metric},
\end{eqnarray}
where $h_{\mu\nu}$ satisfies the transverse and traceless conditions $\partial_{\mu}h^{\mu\nu}=0=\eta^{\mu\nu}h_{\mu\nu}$. By substituting the perturbed metric Eq.~\eqref{perturbed metric} into the field equation~\eqref{Rastal eq}, we derive the linear equation for the tensor perturbation:
\begin{eqnarray}
 -\frac{1}{2}\Box^{(4)} h_{\mu\nu}-\frac{1}{2}h_{\mu\nu}''-\frac{3}{2}A'h_{\mu\nu}'-\left[8\lambda(3A'^2+2A'')\right]h_{\mu\nu}=0, \label{mainequation}\nonumber\\
\end{eqnarray}
where $\Box^{(4)}=\eta^{\alpha\beta}\partial_{\alpha}\partial_{\beta}$. The perturbation $h_{\mu\nu}$ can be decomposed as~\cite{Seahra:2005iq}
\begin{equation}
	h_{\mu\nu}=e^{-\frac{3}{2}A(z)}\Phi(t,z)e^{-i a_{j}x^{j}}\epsilon_{\mu\nu}, ~~~~\epsilon_{\mu\nu}=\text{constant}.\label{decomposition1}
\end{equation}
Substituting the above decomposition (\ref{decomposition1}) into Eq.~(\ref{mainequation}) yields a one-dimensional wave equation for $\Phi(t,z)$
\begin{equation}
	-\partial_{t}^{2}\Phi+\partial_{z}^{2}\Phi-U(z)\Phi-a^{2}\Phi=0, \label{evolutionequation}
\end{equation}
where
\begin{eqnarray}
	U(z)=\left(1-\frac{64}{3}\lambda\right)\left(\frac{3}{2}A''+\frac{9}{4}A'^{2}\right) \label{effectivepotential}
\end{eqnarray}
is the effective potential and the constant $a$ arises from the separation of variables. Moreover, separability implies that the function $\Phi(t,z)$ can be decomposed as
\begin{eqnarray}
	\Phi(t,z)=e^{-i\omega t}\phi(z).\label{decomposition2}
\end{eqnarray}
This leads us to derive a Schr\"odinger-like equation for the extra dimensional part $\phi(z)$
\begin{equation}
	-\partial_{z}^{2}\phi(z)+U(z)\phi(z)=m^{2}\phi(z),\label{Schrodingerlikeequation}
\end{equation}
where $m^2=\omega^2-a^2$ is the mass of the KK modes. Unlike the thick brane solution in general relativity, due to the presence of $\lambda$ in the effective potential \eqref{effectivepotential}, this thick brane model does not have a bound zero mode when $\lambda\neq0$. This is similar to the case of GRS brane and DGP brane, where there is also no bound zero mode in these models, but rather a massive quasibound mode recover the four-dimensional effective gravitational potential. Whether similar quasibound modes exist in rastall branes will discuss in the next section.

\section{quasibound and quasinormal modes of rastall brane}
\label{quasibound and quasinormal modes of rastall brane}
\begin{figure}[htbp]
	\centering
	\includegraphics[width=0.35\textwidth]{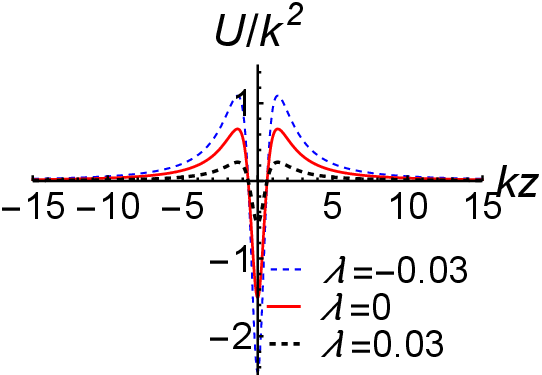}
	\caption{The shapes of the effective potential~(\ref{effectivepotential2}) for different parameter $\lambda$.}\label{effectiveplot}
\end{figure}

In this section, we solve the Schr\"odinger-like equation~\eqref{Schrodingerlikeequation} to obtain the QNMs of the brane. Substituting the warp factor~\eqref{solutionwarpfactor1} into the effective potential~\eqref{effectivepotential}, we derive the specific expression for the effective potential
\begin{eqnarray}
U(z)&=&\frac{k^2(64\lambda-3)\left(5 k^2 z^2-2\right)}{4 \left(k^2 z^2+1\right)^2}.\label{effectivepotential2}
\end{eqnarray}
Plot of the above potential is shown in Fig.~\ref{effectiveplot}. It can be seen that the height of the effective potential decreases with $\lambda$. 

Reference~\cite{Zhong:2022wlw} points out that when $\lambda$ is not equal to 0, the thick brane does not have a bound zero mode. When $\lambda<0$, the thick brane has tachyonic modes and is unstable. Now, we examine in detail the effect of $\lambda$ by solving the QNMs of the thick brane. We use the AIM and the direct integration method to solve the QNMs of the thick brane, and compare the results with the numerical evolution. Since the effective potential $U(z)$ is volcano-like potential and $U(z)\rightarrow0$ as $z\rightarrow\pm \infty$, the reasonable boundary conditions for Eq.~\eqref{Schrodingerlikeequation} are
\begin{equation}
	\label{boundaryconditions}
	\phi(z) \propto \left\{
	\begin{aligned}
		e^{im z}, &~~~~~z\to\infty.& \\
		e^{-im z},  &~~~~~z\to-\infty.&
	\end{aligned}
	\right.
\end{equation}
\subsection{frequency domain}
\label{solveqnms}
\begin{figure}
	\subfigure[~The first QNF]{\label{figlambdarem1}
		\includegraphics[width=0.22\textwidth]{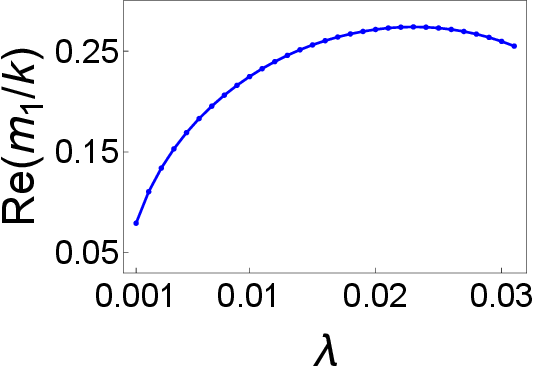}}
	\subfigure[~The first QNF]{\label{figlambdaimm1}
		\includegraphics[width=0.22\textwidth]{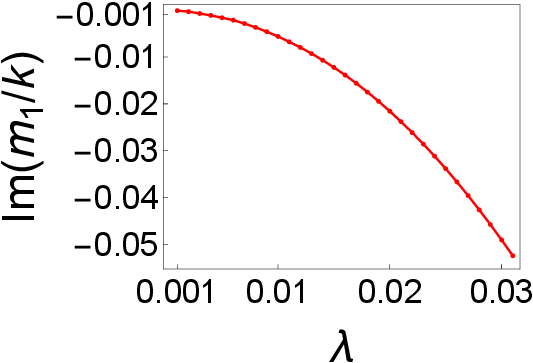}}
	\subfigure[~The second QNF]{\label{figlambdarem2}
		\includegraphics[width=0.22\textwidth]{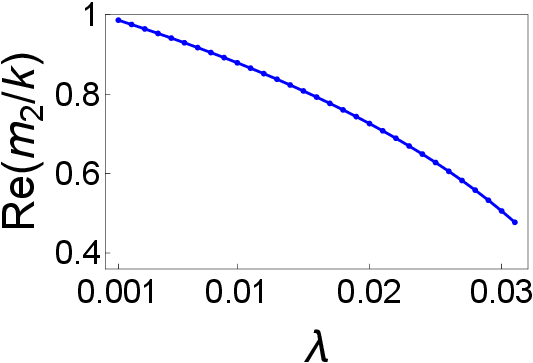}}
	\subfigure[~The second QNF]{\label{figlambdaimm2}
		\includegraphics[width=0.22\textwidth]{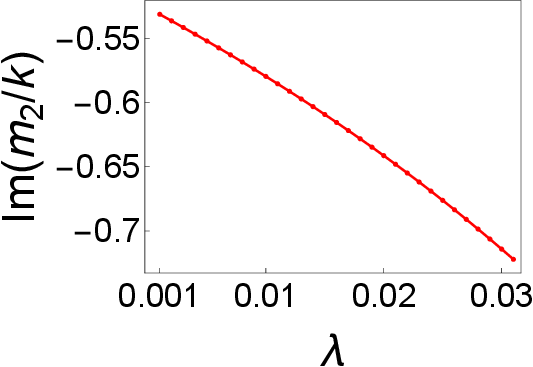}}
	\subfigure[~The third QNF]{\label{figlambdarem3}
		\includegraphics[width=0.22\textwidth]{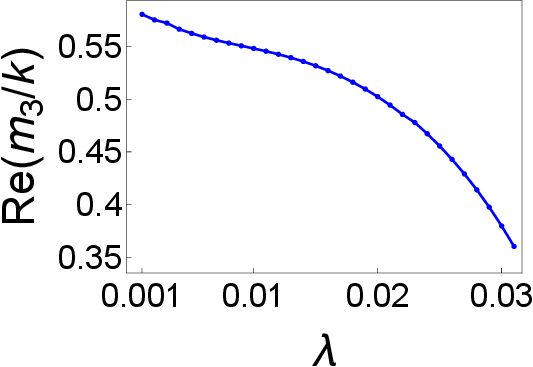}}
	\subfigure[~The third QNF]{\label{figlambdaimm3}
		\includegraphics[width=0.22\textwidth]{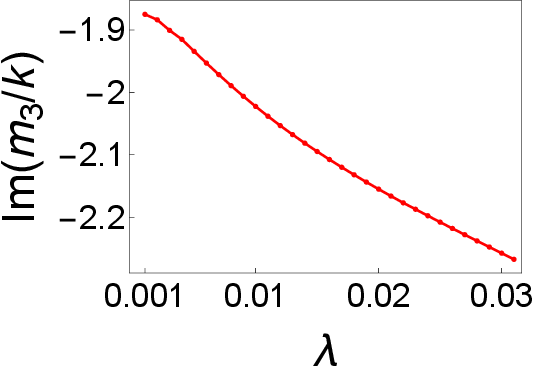}}
	\caption{Left panel: The relation between the real parts of the first three QNFs and the parameter $\lambda$ for $\lambda>0$. Right panel: The relation between the imaginary parts of the first three QNFs and the parameter $\lambda$ for $\lambda>0$.}\label{lambdagt0}
\end{figure}

\begin{figure}[htbp]
	\centering
	\includegraphics[width=0.35\textwidth]{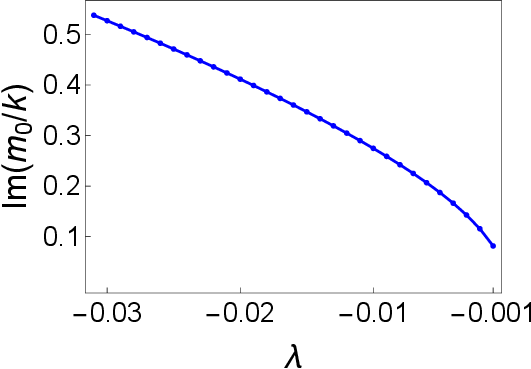}
	\caption{The relation between the imaginary parts of the non-stable mode and the parameter $\lambda$. }\label{figlambdatymode}
\end{figure}

\begin{figure}
	\subfigure[~The first QNF]{\label{figlambdatymoderem1}
		\includegraphics[width=0.22\textwidth]{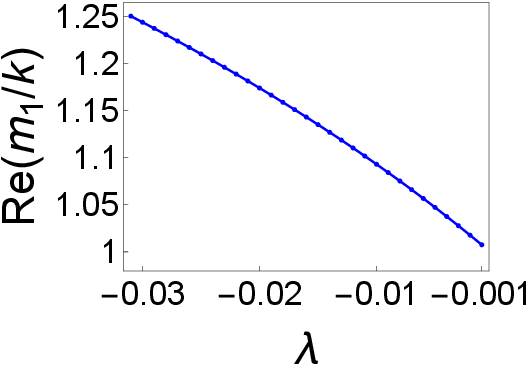}}
	\subfigure[~The first QNF]{\label{figlambdatymodeimm1}
		\includegraphics[width=0.22\textwidth]{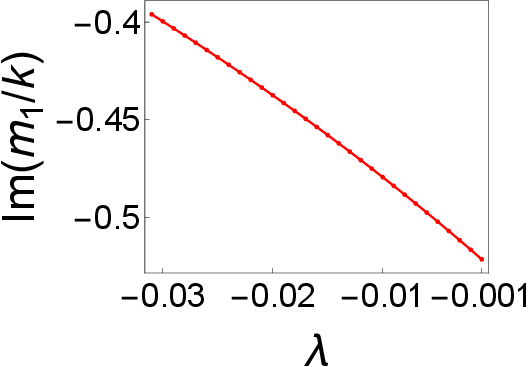}}
	\subfigure[~The second QNF]{\label{figlambdatymoderem2}
		\includegraphics[width=0.22\textwidth]{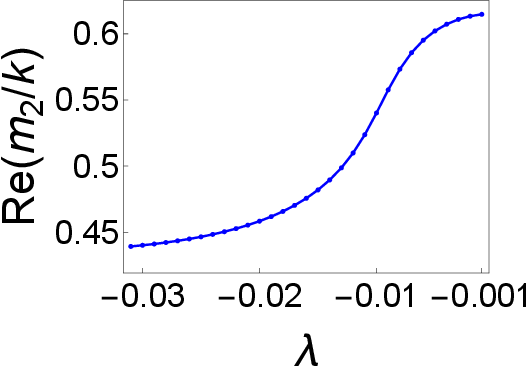}}
	\subfigure[~The second QNF]{\label{figlambdatymodeimm2}
		\includegraphics[width=0.22\textwidth]{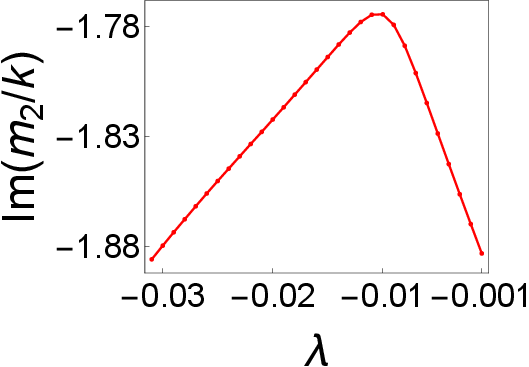}}
	\caption{Left panel: The relation between the real parts of the first two QNFs and the parameter $\lambda$ when $\lambda<0$. Right panel: The relation between the imaginary parts of the first two QNFs and the parameter $\lambda$ when $\lambda<0$.}\label{lambdalt0}
\end{figure}

Let's begin by providing a brief overview of the AIM proposed by Ciftci $et~al.$~\cite{Ciftci:2003As,ciftci:2005co}. This analytical and approximate method is proposed to solving second-order linear differential equations, especially eigenvalue problems. As solving QNMs also is an eigenvalue problem, the AIM is applicable in this context as well. Consider a second-order linear differential equation
\begin{equation}
	y''(x)=h_{0}(x)y'(x)+s_{0}(x)y(x),\label{2orderdiffeq}
\end{equation}
where $h_{0}(x)$ and $s_{0}(x)$ are smooth functions with $h_{0}(x)\neq0$. Differentiating Eq.~\eqref{2orderdiffeq} with respect to $x$, we derive
\begin{eqnarray}
	y'''(x)&=&h_{1}(x)y'(x)+s_{1}(x)y(x),\\
	h_{1}(x)&=&h'_{0}+s_{0}+h_{0}^{2},\\
	s_{1}(x)&=&s'_{0}+s_{0}h_{0}.
\end{eqnarray}
The AIM utilizes the invariant structure of Eq.~\eqref{2orderdiffeq} to find a general solution. Successive Differentiating leads to:
\begin{eqnarray}
	y^{n+1}(x)&=&h_{n-1}(x)y'(x)+s_{n-1}(x)y(x),\\
	y^{n+2}(x)&=&h_{n}(x)y'(x)+s_{n}(x)y(x),	
\end{eqnarray}
where
\begin{eqnarray}
	h_{n}(x)&=&h'_{n-1}+s_{n-1}+h_{0}h_{n-1}, \label{AIMrelation1}\\
	 s_{n}(x)&=&s'_{n-1}+s_{0}h_{n-1}.  \label{AIMrelation2}
\end{eqnarray}
For sufficiently large $n$, the AIM introduce the asymptotic aspect:
\begin{eqnarray}
	\frac{s_{n}(x)}{h_{n}(x)}=\frac{s_{n-1}(x)}{h_{n-1}(x)}=\beta(x).\label{QNMscondition1}
\end{eqnarray}
The quasinormal frequencies (QNFs) are obtained through the ``uantization condition":
\begin{eqnarray}
	s_{n}(x)h_{n-1}(x)-s_{n-1}(x)h_{n}(x)=0.\label{QNMscondition2}
\end{eqnarray}
However, this condition requires derivative operations in each iteration, which is impractical for numerical computations. Cho $et~al.$ introduced an improved AIM version, significantly improving speed and accuracy in numerical calculations~\cite{Cho:2011sf}. The improved AIM expands $s(x)$ and $h(x)$ using Taylor series around a point $\chi$:
\begin{eqnarray}
	h_{n}(x)&=&\sum_{i=0}^{\infty}c_{n}^{i}(x-\chi)^{i},\\
	s_{n}(x)&=&\sum_{i=0}^{\infty}d_{n}^{i}(x-\chi)^{i},
\end{eqnarray}
where $c_{n}^{i}$ and $d_{n}^{i}$ represent the $i$-th Taylor coefficients of $h_{n}$ and $s_{n}$, respectively. The expressions for $c_{n}^{i}$ and $d_{n}^{i}$ are given by:
\begin{eqnarray}
	c_{n}^{i}&=&(i+1)c_{n-1}^{i+1}+d^{i}_{n-1}+\sum_{k=0}^{i}c_{0}^{k}c_{n-1}^{i-k},\\
	d_{n}^{i}&=&(i+1)d_{n-1}^{i+1}+\sum_{k=0}^{i}d_{0}^{k}c_{n-1}^{i-k}.
\end{eqnarray}
Thus, the``quantization condition" ~\eqref{QNMscondition2} transforms into:
\begin{equation}
	d_{n}^{0}c_{n-1}^{0}-d_{n-1}^{0}c_{n}^{0}=0\label{QNMscondition3}.
\end{equation}
Now, we have a set of recursion relations that eliminate the need for derivative operators, simplifying numerical operations. Substituting the warp factor~\eqref{solutionwarpfactor1} into the Schr\"odinger-like equation~\eqref{Schrodingerlikeequation}, we can obtain 
\begin{eqnarray}
	-\partial_{z}^{2}\phi(z)+\left(\frac{k^2(64\lambda-3)\left(5 k^2 z^2-2\right)}{4 \left(k^2 z^2+1\right)^2}-m^{2}\right)\phi(z)=0.\nonumber\\ \label{Schrodingerlikeequationz}
\end{eqnarray}
The asymptotic iterative method requires that the first derivation term of the differential equation is not zero, which is obviously not satisfied by the above equation. So we need to perform the coordinate transformation to obtain the equation containing a first derivative term. Furthermore, the AIM works more effectively on a finite domain. Thus, we introduce the following transformation:
\begin{equation}
	u=\frac{\sqrt{4k^2 z^2+1}-1}{2k z},
\end{equation}
where $-1<u<1$. Now, Eq.~\eqref{Schrodingerlikeequationz} can be rewritten as
 \begin{eqnarray}
	&&\frac{\left(u^2-1\right)^3 \left(\left(u^4-1\right) \phi''(u)+2 u \left(u^2+3\right) \phi'(u)\right)}{\left(u^2+1\right)^3}\nonumber\\
	&&+\left(\frac{m^2}{k^2}-\frac{\left(u^2-1\right)^2 \left(2 u^4-9 u^2+2\right)\left(64\lambda-3\right)}{4\left(u^4-u^2+1\right)^2}\right) \phi(u)=0,\nonumber\\\label{Schrodingerlikeequationu1}
\end{eqnarray}
And the boundary conditions~(\ref{boundaryconditions}) become
\begin{equation}
	\label{transformboundaryconditions}
	\phi(u)\sim\left\{
	\begin{aligned}
		e^{-\frac{i m/k }{2 u-2}}, &~~~ u\to 1& \\
		e^{\frac{i m/k }{2 u+2}}, &~~~ u\to -1&
	\end{aligned}.
	\right.
\end{equation}
Further rewrite $\phi(u)$ as
\begin{eqnarray}
	\phi(u)=\psi (u) e^{-\frac{i m/k }{2 u-2}} e^{\frac{i m/k }{2 u+2}}.\label{boundarysolutions}
\end{eqnarray}
Substituting the above expression into Eq.~\eqref{Schrodingerlikeequationu1} results in:
\begin{equation}
	\psi''(u)=\lambda_{0}(u)\psi'(x)+s_{0}(u)\psi(u),\label{2orderdiffeq_psi}
\end{equation}
where
\begin{eqnarray}
	h_{0}(u)&=&-\frac{2 u \left(u^4+2 i \left(u^2+1\right) \frac{m}{k} +2 u^2-3\right)}{\left(u^2-1\right)^2 \left(u^2+1\right)},\label{lambda0}\\
	s_{0}(u)&=&\frac{1}{4 \left(u^2+1\right) \left(u^6-2 u^4+2 u^2-1\right)^2}\nonumber\\
 &&\times\Bigg[(64 {\lambda}-3) \left(2 u^4-9 u^2+2\right) \left(u^2+1\right)^3\nonumber\\
 &&-4 \left(u^4-u^2+1\right)^2 \left(u^2+1\right) \frac{m^2}{k^2}\nonumber\\
 &&+8 i \left(u^2-1\right) \left(u^4-u^2+1\right)^2 \frac{m}{k}\Bigg].\label{s0}
\end{eqnarray}
With Eq.~\eqref{2orderdiffeq_psi}, we can use the improved AIM to solve the QNMs of the thick brane. After selecting the expansion point $u_{0}=0$, we obtain the quasinormal modes of the thick brane by ``quantization conditions''~\eqref{QNMscondition3}. We investigate the influence of parameter $a$ on the quasinormal modes of the thick brane in two cases: $\lambda>0$ and $\lambda<0$. For the case of $\lambda>0$, the influence of the parameter $\lambda$ on the first three QNFs of the thick brane is shown in Fig.~\ref{lambdagt0}. It can be seen that, the real parts of the second and third QNFs increase with $\lambda$, while the imaginary parts of the first three QNFs decrease with $\lambda$. The behavior of the real part of the first QNF is slightly different: it initially  increases with $\lambda$ and then decreases. It is worth noting that when $\lambda=0$, which corresponds to the case of general relativity, the frequency of the mode with the smallest absolute value of the imaginary part is $\frac{m_1}{k}=0.997018-0.526362i$~\cite{Tan:2022vfe}. However, for $\lambda>0$, there is a quasinormal mode with both real part and absolute value of the imaginary part having magnitudes much smaller than those in the case of general relativity. This is because in the case of general relativity, there exists a bound zero mode, while for $\lambda>0$, there are no bound zero modes in thick brane. Instead, the zero mode is replaced by a long-lived QNM, which is known as a quasibound mode. We propose that this long-lived mode may be crucial for recovering the four-dimensional Newtonian potential on the brane. Similarly, for the case of $\lambda<0$, the influence of the parameter $\lambda$ on the QNFs is shown in Figs.~\ref{figlambdatymode} and~\ref{lambdalt0}. It can be seen from Fig.~\ref{figlambdatymode}, when $\lambda<0$, there is an unstable mode with real part zero and imaginary part greater than zero, that is, a tachyonic mode. The imaginary part of the tachyonic mode decreases with $\lambda$. Thus, the thick brane has tachyonic instability when $\lambda<0$, and this instability decreases with $\lambda$. When $\lambda=0$, this unstable mode becomes a stable bound zero mode. In addition, the QNMs of the thick brane still exist, the real and imaginary parts of the first QNFs decrease with $\lambda$, while real part of the second QNF increases with $\lambda$. The imaginary part of the second QNF first increases with $\lambda$, then decreases, which can be seen from Fig.~\ref{lambdalt0}. In order to verify the accuracy of the above results, we also solved the QNFs of the thick brane by the direct integration method, which are shown in Tab.~\ref{tab1}. The results obtained through several methods are highly consistent within the range of numerical errors, which enhances the credibility of our results. 
On the other hand, in order to more intuitively show the behavior of the long-lived QNM when $\lambda>0$ and the unstable tachyonic mode when $\lambda<0$, we will analyze the time evolution of the initial wave packet on the brane. By combining time-domain and frequency-domain analyses, we will have a clearer understanding of the QNMs of the Rastall brane.

\begin{table*}[htbp]
	\begin{tabular}{|c|c|c|c|c|}
		\hline
		$\;\;\lambda\;\;$  &
		$\;\;n\;\;$  &
		$\;\;\text{Asymptotic iteration method}\;\;$  &
		$\;\;\;\;\;\;\;\;\text{Direct integration method}\;\;\;\;\;\;\;$ 	&
		$\;\;\;\;\;\;\;\;\text{Time evolution}\;\;\;\;\;\;\;$\\
		\hline
		~  &~   &~~~~$\text{Re}(m/k)$  ~~  $\text{Im}(m/k)~~$  &$~~~~~~\text{Re}(m/k)$ ~~ $\text{Im}(m/k)~~$  &$~~~~~~\text{Re}(m/k)$ ~~ $\text{Im}(m/k)~~$     \\
		0.001  &1   &0.079122~~ -0.000119          &~~0.079116~~ -0.000114                &~~0.079123~~ -0.000117     \\
		       &2   &0.986349~~ -0.531389          &~~0.986349~~ -0.531389                &~~1.737690~~ -0.305138 \\		
		0.005  &1   &0.169163~~ -0.001633          &~~0.169150~~ -0.001603                &~~2.355480~~ -0.153401  \\
	           &2   &0.941348~~ -0.552146          &~~0.941348~~ -0.552146                &~~0.940614~~ -0.561521 \\
		0.01~~ &1   &0.224733~~ -0.005624          &~~0.224758~~~-0.005630                &~~0.225071~~~-0.005749 \\
	           &2   &0.879014~~ -0.579726          &~~0.879015~~~-0.579724                &~~0.865270~~~-0.579992 \\
		0~~~~~~~&1  &0.997018~~ -0.526362         &~~0.997020~~  -0.526365                &~~0.991170~~  -0.527546 \\
		-0.001 &1   &0~~~~~~~~~~~~~ 0.081490          &~~1.2$\times10^{-16}$~ 0.080848    &~0~~~~~~~~~~~~    0.083518\\	
		       &2   &1.007486~~ -0.521405          &~~1.007470~~  -0.002554               &~~1.008950~~  -0.526168 \\
		-0.005 &1   &0~~~~~~~~~~~~~ 0.187317        &~~~1.3$\times10^{-17}$~~ 0.187292    &~0~~~~~~~~~~~~ 0.187302 \\
	           &2   &1.047326~~ -0.502166          &~~1.047320~~ -0.502166                &~~1.048877~~ -0.504745  \\
     	-0.01~~&1   &0~~~~~~~~~~~~~ 0.274745          &~~~1.5$\times10^{-16}$~~ 0.274742  &~0~~~~~~~~~~~~ 0.274742 \\
	           &2   &1.093186~~ -0.479369          &~~1.093180~~ -0.479369                &~~1.094242~~ -0.480268 \\
		\hline
	\end{tabular}
	\caption{The first two QNFs using the AIM, the direct integration method, and the time evolution.\label{tab1}}
\end{table*}

\subsection{Time domain}
\label{Time domain}
\begin{figure*}
	\subfigure[~$\lambda=0.001$]{\label{figevenqnms11000}
		\includegraphics[width=0.3\textwidth]{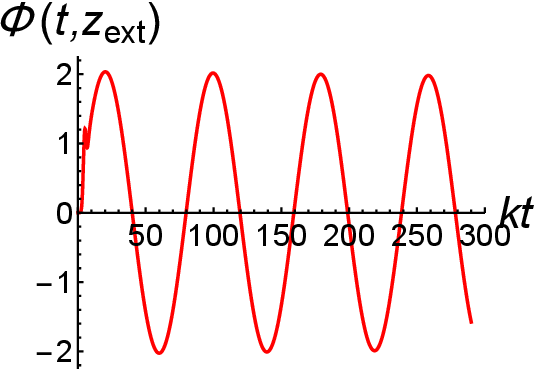}}
	\subfigure[~$\lambda=0.005$]{\label{figevenqnms51000}
		\includegraphics[width=0.3\textwidth]{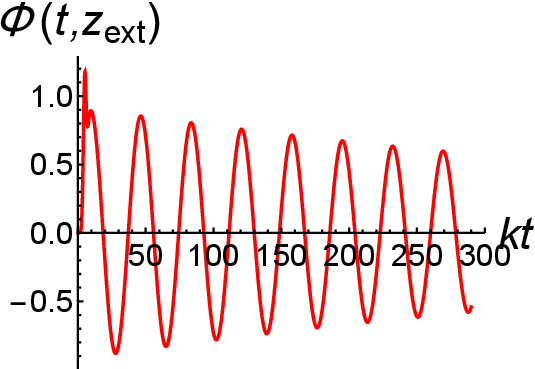}}
	\subfigure[~$\lambda=0.01$]{\label{figevenqnms101000}
		\includegraphics[width=0.3\textwidth]{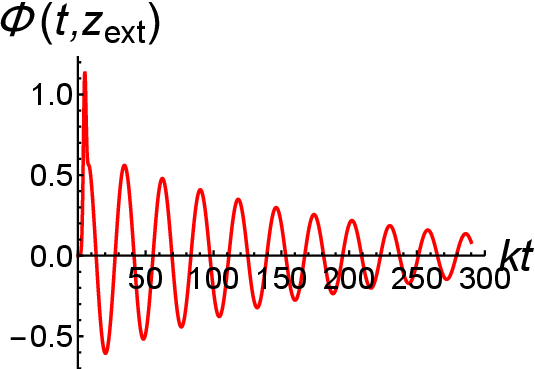}}
			\subfigure[~$\lambda=-0.001$]{\label{figevenqnmsn11000}
			\includegraphics[width=0.3\textwidth]{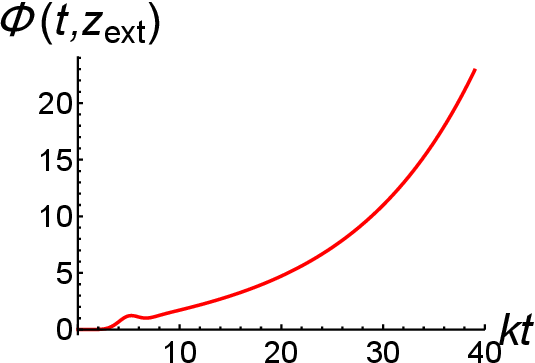}}
		\subfigure[~$\lambda=-0.005$]{\label{figevenqnmsn51000}
			\includegraphics[width=0.3\textwidth]{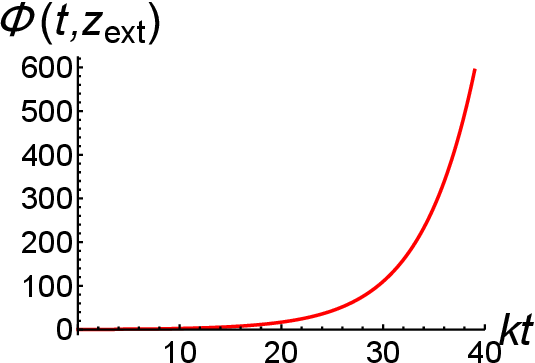}}
		\subfigure[~$\lambda=-0.01$]{\label{figevenqnmsn101000}
			\includegraphics[width=0.3\textwidth]{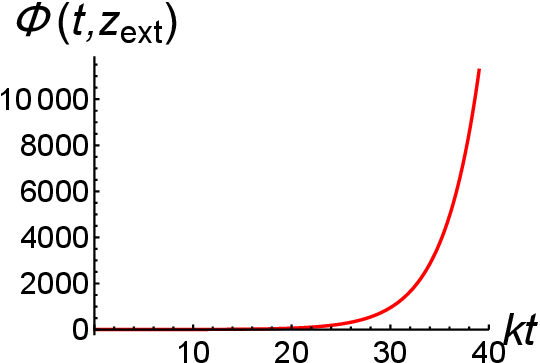}}
	\caption{Upper panel: Time evolution of the Gaussian wave packets with different $\lambda$ at $kz_{\text{ext}}=0$ when $\lambda>0$. Lower panel: Time evolution of the Gaussian wave packets with different $\lambda$ at $kz_{\text{ext}}=0$ when $\lambda<0$.}\label{waveformeven1}
\end{figure*}

\begin{figure*}
	\subfigure[~$\lambda=0.001$]{\label{figlogoddqnms11000}
		\includegraphics[width=0.3\textwidth]{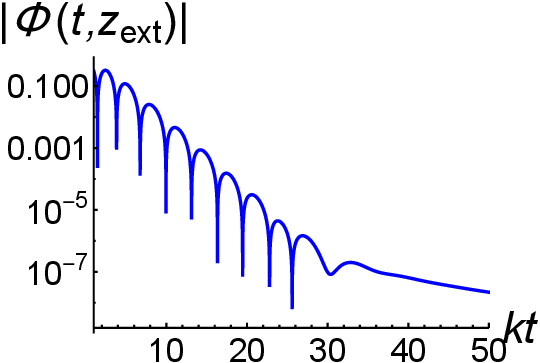}}
	\subfigure[~$\lambda=0.005$]{\label{figlogoddqnms51000}
		\includegraphics[width=0.3\textwidth]{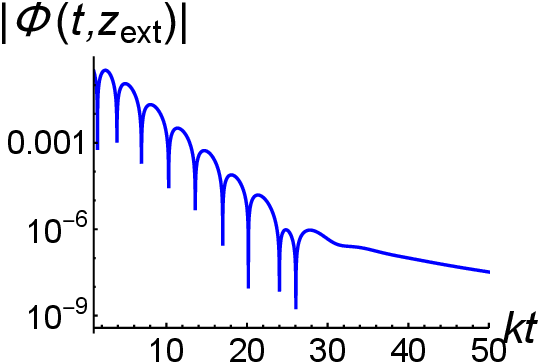}}
	\subfigure[~$\lambda=0.01$]{\label{figlogoddqnms101000}
		\includegraphics[width=0.3\textwidth]{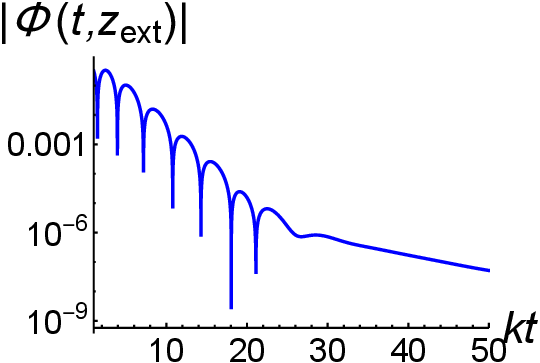}}
	\subfigure[~$\lambda=-0.001$]{\label{figlogevenqnmsn11000}
		\includegraphics[width=0.3\textwidth]{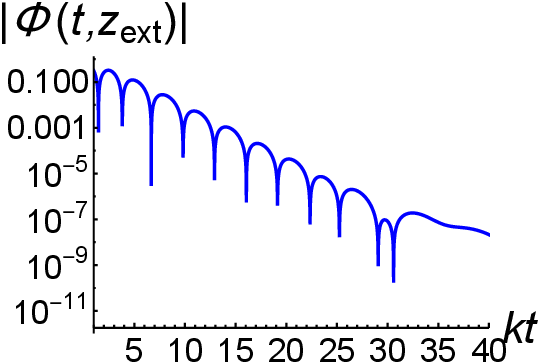}}
	\subfigure[~$\lambda=-0.005$]{\label{figlogoddqnmsn51000}
		\includegraphics[width=0.3\textwidth]{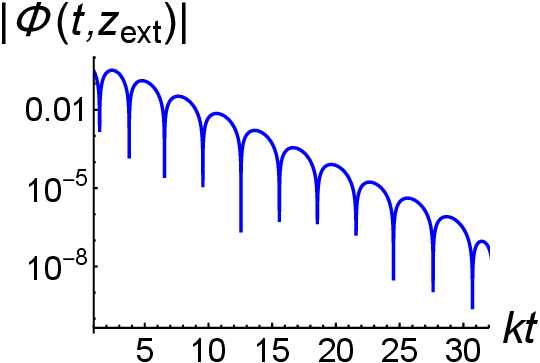}}
	\subfigure[~$\lambda=-0.01$]{\label{figlogoddqnmsn101000}
		\includegraphics[width=0.3\textwidth]{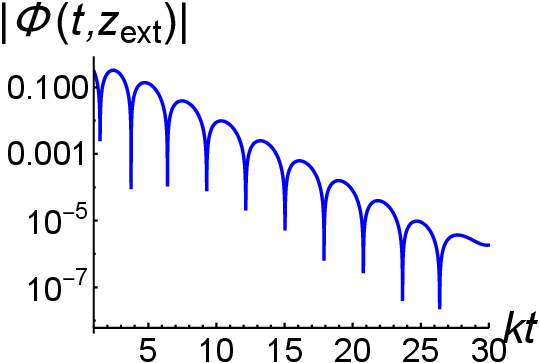}}
	\caption{Upper panel: Time evolution of the odd wave packets with different $\lambda$ at $kz_{\text{ext}}=0$ when $\lambda>0$. Lower panel: Time evolution of the odd wave packets with different $\lambda$ at $kz_{\text{ext}}=1$ when $\lambda<0$.}\label{waveformodd1}
\end{figure*}

Now we consider the evolution of an initial wave packet on the thick brane. We use the following coordinates: $u = t + z$, $v = t - z$. In this case, Eq.~\eqref{evolutionequation} becomes:
\begin{eqnarray}
	\left(4\frac{\partial^{2}}{\partial u\partial v}+U+a^{2}\right)\Phi(u,v)=0. \label{uvevolutionequation}
\end{eqnarray}
Since the thick brane is symmetric, the KK modes are either odd or even. Obviously, different initial wave packets will excite different modes, so we consider two types of initial wave packets, the first one is a Gaussian pulse:
\begin{eqnarray}
	\Phi(0,v)=e^{\frac{-(v-v_{c})^{2}}{2\sigma^{2}}}, ~~~\Phi(u,0)=e^{\frac{-v_{c}^{2}}{2\sigma^{2}}},\label{gausspulseinitialwavepackage}
\end{eqnarray}
and the second being an odd initial wave packet:
\begin{eqnarray}
	\Phi(0,v)=\sin\left(\frac{kv}{2}\right)e^{\frac{-k^2v^{2}}{4}}, \\
	\Phi(u,0)=\sin\left(\frac{ku}{2}\right)e^{\frac{-k^2u^{2}}{4}}.\label{oddinitialwavepackage}
\end{eqnarray}
First, we consider the evolution of the Gaussian wave packet with $k\sigma=1$ and $kv_{c}=5$ on the brane. Note that in this paper we only consider the case of $a=0$. We plot the evolution waveforms of the Gaussian pulses for different $\lambda$ in Fig.~\ref{waveformeven1}.  From Figs.~\ref{figevenqnms11000}, \ref{figevenqnms51000},and \ref{figevenqnms101000}, it can be seen that the evolution waveform of the initial wave packet is a damped sinusoids. The decay rate of the wave packet increases with the parameter $\lambda$. In other words, the further away from the general relativity scenario, the shorter the lifetime of the first QNM. When  $\lambda\rightarrow0$, a non-decaying bound zero mode appears. In contrast, the situation is significantly different when $\lambda<0$. As seen from Figs.~\ref{figevenqnmsn11000}, \ref{figevenqnmsn51000}, and~\ref{figevenqnmsn101000}, the evolving waveform exhibits exponential growth with time, demonstrating tachyonic instability. The growth rate of the tachyonic mode decreases with the parameter $\lambda$, meaning that although the thick brane is unstable for $\lambda<0$, this instability decreases with $\lambda$. Furthermore, we extracted the frequencies of these waveforms through waveform fitting, and the results are presented in Tab.~\ref{tab1}. It can be seen that the extracted data aligns with the frequency domain analysis results. Then, we consider the evolution of the odd wave packet. The waveforms for different $\lambda$ are shown in Fig.~\ref{waveformodd1}. It can be seen that regardless of whether $\lambda>0$ or $\lambda<0$, the amplitude of the waveform rapidly oscillates and decays, consistent with the case of general relativity~\cite{Tan:2022vfe}. Similarly, we also extracted the frequencies of these QNMs, and the results are shown in Tab.~\ref{tab1}. Combining the results from the frequency domain and time domain, we can confirm that when $\lambda>0$, the bound zero mode transitions into a decaying long-lived QNM, whereas when $\lambda<0$, the zero mode transitions into a growing tachyonic mode. This shows that the thick brane is unstable in this case and is not a reasonable thick brane model.

A suitable model for a thick brane should be stable and capable of recovering the effective gravitational potential on the brane. From this perspective, the Rastall thick brane appears to be unsuccessful, especially for the case of $\lambda<0$. However, for $\lambda>0$, the presence of the long-lived QNM makes the situation interesting. Despite the absence of the bound zero mode, the effective gravitational potential can be recovered within a certain range by a quasi-bound gravitational KK mode. This is also the core idea of the DGP model and the GRS model: within the screening radius, the gravitational potential on the brane is approximately four-dimensional, but for scales larger than the lifetime of the long-lived KK mode, the five-dimensional gravitational potential will be reproduced. According to Refs.~\cite{Dvali:2000hr,Gregory:2000jc,Csaki:2000pp}, when there is no bound zero mode but long-lived KK modes with a small KK mass $m$ exist, the Newtonian potential on the brane is
\begin{eqnarray}
U(r)\sim&&\frac{1}{r}\Big[\text{Ci}\left(\frac{2r}{r_{c}}\right)\sin\left(\frac{2r}{r_{c}}\right)\nonumber\\
&&+\frac{1}{2}\cos\left(\frac{2r}{r_{c}}\right)\left(\pi-2\text{Si}\left(\frac{2r}{r_{c}}\right)\right)\Big],\label{Newtonpotential}
\end{eqnarray}
where $\text{Ci(t)}=-\int_{t}^{\infty}\frac{\cos(u)}{u}du$ and $\text{Si(t)}=-\int_{t}^{\infty}\frac{\sin(u)}{u}du$ are the cosine and sine integrals, respectively. And $r_{c}\sim\frac{1}{Im(m_{1})}$ is the screening radius. At $r\ll r_{c}$, the four-dimensional gravitational potential is restored, $U(r)\sim\frac{1}{r}$. But for the case of $r\gg r_{c}$, the gravitational potential is $U(r)\sim\frac{r_{c}}{r^2}$. Clearly, the screening radius is closely related to the imaginary part of the long-lived QNM. In the case of the Rastall brane, when the parameter $\lambda$ is small enough, both the real and imaginary parts of the long-lived modes are extremely small, and the screening radius is also greatly enlarged. In this way, the inverse square law of gravity is satisfied on a large scale. Therefore, perhaps we do not need bound zero modes to meet current observations.

\section{Conclusion and discussion}
\label{Conclusion}
In this paper, we investigated the QNMs of the Rastall thick brane. Using the asymptotic iteration method, direct integration method, and numerical evolution, we obtained the frequencies and waveforms of the QNMs for the Rastall thick brane. The results from these three methods were consistent. It shows that depending on the choice of parameter $\lambda$, the thick brane exhibited the long-lived quasibound modes, the QNMs, and the tachyonic mode. When the parameter $\lambda\neq0$, the thick brane did not have bound zero modes, but under the condition where $\lambda$ was sufficiently close to $0_{+}$, the existence of long-lived gravitational QNMs allowed for a substantial restoration of the four-dimensional Newtonian potential on a large scale, which was not contradictory to current observations. This brings some enlightenment for the study of quasi-localization of gravity in the thick brane model.

We reviewed the solutions of the Rastall thick brane and its transverse-traceless tensor perturbation, obtaining the main equations~\eqref{evolutionequation} and \eqref{Schrodingerlikeequation}. Based on these equations, we investigated the properties of QNMs of the thick brane from both frequency domain and time domain. Using the AIM and direct integration method, we solved the QNFs of the thick brane, and the results were presented in Figs.~\ref{lambdagt0},~\ref{figlambdatymode},~\ref{lambdalt0} and Tab.~\ref{tab1}. It was observed that when $\lambda>0$, the thick brane exhibit a long-lived QNM with an extremely small imaginary part, while when $\lambda<0$, an unstable tachyonic mode appeared, with neither case having bound zero mode. This is the most direct difference from the general relativity thick brane. Additionally, the parameter $\lambda$ have a certain influence on the QNFs of the thick brane. After obtaining the frequency domain results, we further used the numerical evolution method to obtain the evolving waveforms of the KK modes and extracted the frequencies of the corresponding modes, which were presented in Figs.~\ref{waveformeven1},~\ref{waveformodd1}, and Tab.~\ref{tab1}. The frequencies extracted using numerical evolution are consistent with the frequency domain results, indicating the reliability of our results. Finally, we discussed the possibility of restoring the four-dimensional effective Newtonian potential on the Rastall thick brane and found that for $\lambda\gtrsim0$, that is, when deviating only slightly from general relativity, the long-lived mode can replace the bound zero mode to generate a four-dimensional effective Newtonian potential within a large scale. This provides new research insights into the localization of gravity on the thick branes.

There are several ways in which our work could be improved. For example, we could investigate the QNMs of scalar perturbation of the thick brane, quasinormal modes of some non-flat or non-symmetric thick branes are also worth studying.

\section*{Acknowledgements}
This work was supported by the National Natural Science Foundation of China (Grants No. 12347111, No. 12205129, and No. 12305061), the China Postdoctoral Science Foundation (Grants No. 2021M701529 and No. 2023M741148), and the Natural Science Foundation of Hunan Province, China (Grant No. 2022JJ40033).


\begin{thebibliography}{99}
	
	
	
\bibitem{kaluza:1921un}
T.~Kaluza, \emph{Zum unit{\"a}tsproblem der physik}, {\emph{Sitzungsber.
		Preuss. Akad. Wiss. Berlin (Math. Phys.)} {\bfseries 27} (1921) 966}.

\bibitem{Klein:1926tv}
O.~Klein, \emph{{Quantum Theory and Five-Dimensional Theory of Relativity. (In
	German and English)}}, {\emph{Z.
		Phys.} {\bfseries 37} (1926) 895}.	
	
	
\bibitem{Arkani-Hamed:1998jmv}
N.~Arkani-Hamed, S.~Dimopoulos, and G.~R.~Dvali,
\emph{{The Hierarchy problem and new dimensions at a millimeter}},
{\emph{Phys. Lett. B}  {\bfseries429} , 263 (1998)},
[{{\ttfamily arXiv:hep-ph/9803315}}].


\bibitem{Antoniadis:1998ig}
I.~Antoniadis, N.~Arkani-Hamed, S.~Dimopoulos, and G.~R.~Dvali,
\emph{{New dimensions at a millimeter to a Fermi and superstrings at a TeV,}}
{\emph{ Phys. Lett. B}
	{\bfseries 436}, 257 (1998)},
[{{\ttfamily arXiv:hep-ph/9804398}}].

\bibitem{Randall:1999ee}
L.~Randall and R.~Sundrum, \emph{{A Large mass hierarchy from a small extra dimension}},
{\emph{Phys. Rev. Lett.}
	{\bfseries 83},  3370 (1999)},
[{{\ttfamily arXiv:hep-ph/9905221}}].

\bibitem{Randall:1999vf}
L.~Randall and R.~Sundrum, \emph{{An Alternative to compactification}},
{\emph{Phys. Rev. Lett.}
	{\bfseries 83},  4690 (1999)},
[{{\ttfamily arXiv:hep-th/9906064}}].





\bibitem{Shiromizu:1999wj}
T.~Shiromizu, K.~Maeda, and M.~Sasaki,
{\emph{The Einstein equation on the 3-brane world}},
{\emph{Phys. Rev. D} {\bfseries 62},  024012 (2000)},
[{{\ttfamily arXiv:gr-qc/9910076}}].


\bibitem{Tanaka:2002rb}
T.~Tanaka,
{\emph{Classical black hole evaporation in Randall-Sundrum infinite brane world}},
{\emph{Prog. Theor. Phys. Suppl.} {\bfseries 148},  307 (2003)},
[{{\ttfamily arXiv:gr-qc/0203082}}].

\bibitem{Gregory:2008rf}
R.~Gregory,
{\emph{Braneworld black holes}},
{\emph{Lect. Notes Phys.}  {\bfseries 769},  259 (2009)},
[{{\ttfamily arXiv:0804.2595}}].


\bibitem{Jaman:2018ucm}
N.~Jaman and K.~Myrzakulov,
{\emph{Braneworld inflation with an effective $\alpha$-attractor potential}},
{\emph{Phys. Rev. D} {\bfseries 99}, 103523 (2019)},
[{{\ttfamily arXiv:1807.07443}}].


\bibitem{Adhikari:2020xcg}
R.~Adhikari, M.~R.~Gangopadhyay, and Yogesh,
{\emph{Power Law Plateau Inflation Potential In The RS $II$ Braneworld Evading Swampland Conjecture}},
{\emph{Eur. Phys. J. C} {\bfseries 80}, 899 (2020)},
[{{\ttfamily arXiv:2002.07061}}].


\bibitem{Bhattacharya:2021jrn}
A.~Bhattacharya, A.~Bhattacharyya, P.~Nandy, and A.~K.~Patra,
{\emph{Islands and complexity of eternal black hole and radiation subsystems for a doubly holographic model}},
{\emph{JHEP} {\bfseries 05}, 135 (2021)},
[{{\ttfamily arXiv:2103.15852}}].

\bibitem{Geng:2020fxl}
H.~Geng, A.~Karch, C.~Perez-Pardavila, S.~Raju, L.~Randall, M.~Riojas, and S.~Shashi,
{\emph{Information Transfer with a Gravitating Bath}},
{\emph{SciPost Phys.} {\bfseries 10} 103 (2021)} , 
[{{\ttfamily arXiv:2012.04671}}].



\bibitem{Geng:2021iyq}
H.~Geng, S.~L\"ust, R.~K.~Mishra, and D.~Wakeham,
{\emph{Holographic BCFTs and Communicating Black Holes}},
{\emph{JHEP} {\bfseries 08} 003 (2021)}, 
[{{\ttfamily arXiv:2104.07039}}].

\bibitem{DeWolfe:1999cp}
O.~DeWolfe, D.~Z.~Freedman, S.~S.~Gubser, and A.~Karch,
\emph{{Modeling the fifth-dimension with scalars and gravity}},
{\emph{Phys. Rev. D} {\bfseries 62}, 046008 (2000)},
[{{\ttfamily arXiv:hep-th/9909134}}].


\bibitem{Gremm:1999pj}
M.~Gremm,
\emph{{Four-dimensional gravity on a thick domain wall}},
{\emph{Phys. Lett. B} {\bfseries 478}, 434 (2000)},
[{{\ttfamily arXiv:hep-th/9912060}}].


\bibitem{Csaki:2000fc}
C.~Csaki, J.~Erlich, T.~J.~Hollowood, and Y.~Shirman,
\emph{{Universal aspects of gravity localized on thick branes}},
{\emph{Nucl. Phys. B} {\bfseries 581}, 309 (2000)},
[{{\ttfamily arXiv:hep-th/0001033}}].



\bibitem{Afonso:2007gc}
V.~I.~Afonso, D.~Bazeia, R.~Menezes, and A.~Y.~Petrov,
\emph{{f(R)-Brane}},
{\emph{Phys. Lett. B} {\bfseries 658}, 71 (2007)},
[{{\ttfamily arXiv:0710.3790}}].




\bibitem{Dzhunushaliev:2010fqo}
V.~Dzhunushaliev and V.~Folomeev,
\emph{{Spinor brane}},
{\emph{Gen. Rel. Grav.} {\bfseries 43}, 1253  (2011)},
[{{\ttfamily arXiv:0909.2741}}].



\bibitem{Dzhunushaliev:2011mm}
V.~Dzhunushaliev and V.~Folomeev,
\emph{{Thick brane solutions supported by two spinor fields}},
{\emph{Gen. Rel. Grav.} {\bfseries 44}, 253  (2012)},
[{{\ttfamily arXiv:1104.2733}}].


\bibitem{Geng:2015kvs}
W.-J.~Geng and H.~Lu,
\emph{{Einstein-Vector Gravity, Emerging Gauge Symmetry and de Sitter Bounce}},
{\emph{Phys. Rev. D} {\bfseries 93}, 044035  (2016)},
[{{\ttfamily arXiv:1511.03681}}].



\bibitem{Melfo2006}
A.~Melfo, N.~Pantoja, and J.~D. Tempo, {\it {Fermion localization on thick branes}},  {\em Phys. Rev. D} {\bf 73},  044033 (2006), [{{\tt arXiv:hep-th/0601161}}].

\bibitem{Almeida2009}
C.~A. Almeida, R.~Casana, M.~M. Ferreira, and A.~R. Gomes, {\it {Fermion localization and resonances on two-field thick branes}},  {\em Phys. Rev. D} {\bf 79},  125022 (2009), [{{\tt arXiv:0901.3543}}].

\bibitem{Zhao2010}
Z.-H. Zhao, Y.-X. Liu, and H.-T. Li, {\it {Fermion localization on asymmetric two-field thick branes}},  {\em Class. Quantum Gravity} {\bf 27},  185001 (2010), [{{\tt arXiv:0911.2572}}].

\bibitem{Chumbes2011}
A.~E.~R. Chumbes, A.~E.~O. Vasquez, and M.~B. Hott, {\it {Fermion localization on a split brane}},  {\em Phys. Rev. D} {\bf 83},  105010 (2011), [{{\tt arXiv:1012.1480}}].

\bibitem{Liu2011}
Y.-X. Liu, Y.~Zhong, Z.-H. Zhao, and H.-T. Li,
{\it {Domain wall brane in squared curvature gravity}},  
{\em J. High Energy Phys.} {\bf 2011}, 135 (2011), 
[{{\tt arXiv:1104.3188v2}}].



\bibitem{Bazeia:2013uva}
D.~Bazeia, A.~S.~Lob\~ao, Jr., R.~Menezes, A.~Y.~Petrov, and A.~J.~da Silva,
{\it {Braneworld solutions for F(R) models with non-constant curvature}},
{\em Phys. Lett. B} {\bf 729}, 127 (2014),
[{{\tt arXiv:1311.6294}}].



\bibitem{Xie2017}
Q.-Y. Xie, H.~Guo, Z.-H. Zhao, Y.-Z. Du, and Y.-P. Zhang, {\it {Spectrum structure of a fermion on Bloch branes with two scalar-fermion couplings}}, {\em Class. Quantum Gravity} {\bf 34},  055007 (2017), [{{\tt arXiv:1510.03345}}].

\bibitem{Gu2017}
B.-M. Gu, Y.-P. Zhang, H.~Yu, and Y.-X. Liu, {\it {Full linear perturbations and localization of gravity on $f(R, T)$ brane}},  {\em Eur. Phys. J. C} {\bf 77}, 115 (2017), [{{\tt arXiv:1606.07169}}].

\bibitem{ZhongYuan2017}
Y.~Zhong and Y.-X. Liu, {\it {Linearization of a warped $f(R)$ theory in the higher-order frame}},  {\em Phys. Rev. D} {\bf 95}, 104060 (2017),
[{{\tt arXiv:1611.08237}}].

\bibitem{ZhongYuan2017b}
Y.~Zhong, K.~Yang, and Y.-X. Liu, {\it {Linearization of a warped $f(R)$ theory in the higher-order frame II: The equation of motion approach}}, {\em Phys. Rev. D} {\bf 97},  044032 (2017), [{{\tt arXiv:1708.03737}}].

\bibitem{Zhou2018}
X.-N. Zhou, Y.-Z. Du, H.~Yu, and Y.-X. Liu,
{\it {Localization of gravitino field on $f(R)$-thick branes}},
{\em Sci. China Physics, Mech. Astron.} {\bf 61}, 110411 (2018),
[{{\tt arXiv:1703.10805}}].



\bibitem{Hendi:2020qkk}
S.~H.~Hendi, N.~Riazi, and S.~N.~Sajadi,
{\it {$Z_2$-symmetric thick brane with a specific warp function}},
{\em Phys. Rev. D} {\bf 102}, 124034 (2020),
[{{\tt arXiv:2011.11093}}].



\bibitem{Xie:2021ayr}
Q.-Y.~Xie, Q.-M.~Fu, T.-T.~Sui, L.~Zhao, and Y.~Zhong,
{\it {First-Order Formalism and Thick Branes in Mimetic Gravity}},
{\em Symmetry} {\bf 13}, 1345 (2021),
[{{\tt arXiv:2102.10251}}].



\bibitem{Moreira:2021uod}
A.~R.~P.~Moreira, F.~C.~E.~Lima, J.~E.~G.~Silva, and C.~A.~S.~Almeida,
{\it {First-order formalism for thick branes in $f(T,{\mathscr {T}})$ gravity}},
{\em Eur. Phys. J. C}   {\bf 81}, 1081 (2021),
[{{\tt arXiv:2107.04142}}].


\bibitem{Xu:2022ori}
N.~Xu, J.~Chen, Y.-P.~Zhang, and Y.-X.~Liu,
{\it {Multi-kink brane in Gauss-Bonnet gravity}},
[{{\tt arXiv:2201.10282}}].



\bibitem{Silva:2022pfd}
J.~E.~G.~Silva, R.~V.~Maluf, G.~J.~Olmo, and C.~A.~S.~Almeida,
{\it {Braneworlds in $f(Q)$ gravity}},
[{{\tt arXiv:2203.05720}}].


\bibitem{Xu:2022gth}
Y.-Q.~Xu and X.-D.~Zhang,
{\it {Tensor Perturbations and Thick Branes in Higher Dimensional Gauss-Bonnet Gravity}},
[{{\tt arXiv:2203.13401}}].

\bibitem{Liu:2017gcn}
Y.-X.~Liu,
{\it {Introduction to Extra Dimensions and Thick Braneworlds}},
[{{\tt arXiv:1707.08541}}].


\bibitem{Ahluwalia:2022ttu}
D.~V.~Ahluwalia, J.~M.~H.~da Silva, C.~Y.~Lee, Y.-X.~Liu, S.~H.~Pereira, and M.~M.~Sorkhi,
\emph{{Mass dimension one fermions: Constructing darkness}},
{\emph{Phys. Rept.}} {\bfseries 967}, 1 (2022),
[{{\tt arXiv:2205.04754}}].



\bibitem{Dzhunushaliev:2009va}
V.~Dzhunushaliev, V.~Folomeev, and M.~Minamitsuji,
\emph{{Thick brane solutions}},
{\emph{Rept. Prog. Phys.}} {\bfseries 73}, 066901  (2010),
[{{\tt arXiv:0904.1775}}].





\bibitem{Dvali:2000hr}
G.~R.~Dvali, G.~Gabadadze, and M.~Porrati,
{\it {4-D gravity on a brane in 5-D Minkowski space}},
{\em Phys. Lett. B} {\bf 485}, 208 (2000),
[{{\tt arXiv:hep-th/0005016}}].



\bibitem{Gregory:2000jc}
R.~Gregory, V.~A.~Rubakov and S.~M.~Sibiryakov,
{\it {Opening up extra dimensions at ultra large scales}},
{\em Phys. Rev. Lett.} {\bf  84}, 5928 (2000),
[{{\tt arXiv:hep-th/0002072}}].



\bibitem{Zhong:2022wlw}
Y.~Zhong, K.~Yang and Y.~X.~Liu,
{\it {Thick brane in Rastall gravity}},
{\em JHEP}  {\bf  09}, 128 (2022),
[{{\tt arXiv:2206.15145}}].



\bibitem{Rastall:1972swe}
P.~Rastall,
{\it {Generalization of the einstein theory}},
{\em Phys. Rev. D} {\em 6}, 3357  (1972).



\bibitem{Parker:1969au}
L.~Parker,
{\it {Quantized fields and particle creation in expanding universes. 1.}},
{\em Phys. Rev.}{\em 183}, 1057 (1969).




\bibitem{Gibbons:1977mu}
G.~W.~Gibbons and S.~W.~Hawking,
{\it {Cosmological Event Horizons, Thermodynamics, and Particle Creation}},
{\em  Phys. Rev. D} {\em 15}, 2738   (1977).


\bibitem{Ford:1986sy}
L.~H.~Ford,
{\it {Gravitational Particle Creation and Inflation}},
{\em  Phys. Rev. D} {\em 35}, 2955  (1987).


\bibitem{BezerradeMello:2014okn}
E.~R. Bezerra~de Mello, J.~C. Fabris, and B.~Hartmann,
\emph{{Abelian\textendash{}Higgs strings in Rastall gravity}},
{\emph{Class. Quant. Grav.} {\bf 32},  085009 (2015)},
	 [{{\tt  arXiv:1407.3849}}].

\bibitem{Heydarzade:2017wxu}
Y.~Heydarzade and F.~Darabi, 
\emph{{Black Hole Solutions Surrounded by Perfect Fluid in Rastall Theory}},
{\emph{Phys. Lett. B}{\bf 771}, 365 (2017)}, 
	[{{\tt  arXiv:1702.07766}}].

\bibitem{Darabi:2017tay}
F.~Darabi, K.~Atazadeh, and Y.~Heydarzade,
 \emph{{Einstein static universe in the Rastall theory of gravity}},
{\emph{Eur. Phys. J. Plus} {\bf 133}, 249  (2018)},
	 [{{\tt arXiv:1710.10429}}].

\bibitem{Xu:2017vse}
Z.~Xu, X.~Hou, X.~Gong, and J.~Wang, 
\emph{{Kerr\textendash{}Newman-AdS black hole surrounded by perfect fluid matter in Rastall gravity}},
{\emph{Eur. Phys. J. C} {\bf  78}, 513  (2018)}, 
[{{\tt	arXiv:1711.04542}}].

\bibitem{Darabi:2017coc}
F.~Darabi, H.~Moradpour, I.~Licata, Y.~Heydarzade, and C.~Corda, 
\emph{{Einstein and Rastall Theories of Gravitation in Comparison}},
{\emph{Eur. Phys. J. C} {\bf  78}, 25  (2018)}, 
[{{\tt  arXiv:1712.09307}}].

\bibitem{Das:2018dzp}
D.~Das, S.~Dutta and, S.~Chakraborty, 
\emph{{Cosmological consequences in the framework of generalized Rastall theory of gravity}},
{\emph{Eur. Phys. J. C} {\bf  78}, 810 (2018)},
 [{{\tt  arXiv:1810.11260}}].

\bibitem{Tang:2019dsk}
M.~Tang, Z.~Xu, and J.~Wang, 
\emph{{Observational constraints on Rastall gravity from rotation curves of low surface brightness galaxies}},
{\emph{Chin. Phys.}{\bf C 44}, 085104  (2020)},
	 [{{\tt arXiv:1903.01034}}].

\bibitem{Li:2019jkv}
R.~Li, J.~Wang, Z.~Xu, and X.~Guo, 
\emph{{Constraining the Rastall parameters in	static space\textendash{}times with galaxy-scale strong gravitational lensing}},
	{\emph{Mon. Not. Roy. Astron. Soc.} {\bf 486}, 2407 (2019)},
[{{\tt  arXiv:1903.08790}}].

\bibitem{Khyllep:2019odd}
W.~Khyllep and J.~Dutta,
 \emph{{Linear growth index of matter perturbations in Rastall gravity}},
{\emph{Phys. Lett.}{\bf B 797}, 134796  (2019)},
	 [{{\tt  arXiv:1907.09221}}].

\bibitem{Ghosh:2021byh}
S.~Ghosh, S.~Dey, A.~Das, A.~Chanda, and B.~C. Paul, 
\emph{{Study of gravastars in Rastall gravity}},
	{\emph{JCAP} {\bf 07}, 004 (2021)}, 
	[{{\tt arXiv:2102.01524}}].

\bibitem{Haghani:2022lsk}
Z.~Haghani and T.~Harko, 
\emph{{Compact stars in the Einstein dark energy model}},
	{\emph{Phys. 	Rev.} {\bf D 105}, 064059   (2022)},
[{{\tt arXiv:2203.05764}}].

\bibitem{Shahidi:2021lxt}
S.~Shahidi, 
\emph{{Cosmological implications of Rastall-f(R) theory}},
{\emph{Phys. Rev.} {\bf	D 104}, 084033  (2021)}  ,
 [{{\tt arXiv:2108.00423}}].


\bibitem{Shabani:2020wja}
H.~Shabani and A.~Hadi~Ziaie,
 \emph{{A connection between Rastall-type and $f(R, T)$ gravities}},
{\emph{EPL} {\bf 129}, 20004 (2020)},
 [{{\tt arXiv:2003.02064}}].

\bibitem{Tan:2022vfe}
Q.~Tan, W.-D.~Guo, and Y.-X.~Liu,
\emph{{Sound from extra dimension: quasinormal modes of a thick brane}},
{\emph{Phys. Rev. D}  {\bf  106}, 044038 (2022)},
[{{\tt arXiv:2205.05255}}].



\bibitem{Tan:2023cra}
Q.~Tan, W.-D.~Guo, Y.-P.~Zhang, and Y.-X.~Liu,
\emph{{Characteristic modes of a thick brane: Resonances and quasinormal modes}},
{\emph{Phys. Rev. D}  {\bf  109}, 2 (2024)},
[{{\tt  arXiv:2304.09363}}].









\bibitem{Berti:2009kk}
E.~Berti, V.~Cardoso, and A.~O.~Starinets,
\emph{{Quasinormal modes of black holes and black branes}},
{\emph{Class. Quant. Grav.} {\bfseries 26},
	163001 (2009)},
[{{\ttfamily arXiv:0905.2975}}].


\bibitem{Kokkotas:1999bd}
K.~D.~Kokkotas and B.~G.~Schmidt,
\emph{{Quasinormal modes of stars and black holes}},
{\emph{Living Rev. Rel.} {\bfseries 2}, 2 (1999)},
[{{\ttfamily  arXiv:gr-qc/9909058}}].


\bibitem{Nollert:1999ji}
H.~P.~Nollert,
\emph{{TOPICAL REVIEW: Quasinormal modes: the characteristic `sound' of black holes and neutron stars}},
{\emph{Class. Quant. Grav.} {\bfseries 16}, R159 (1999)}.


\bibitem{Konoplya:2011qq}
R.~A.~Konoplya and A.~Zhidenko,
\emph{{Quasinormal modes of black holes: From astrophysics to string theory}},
{\emph{ Rev. Mod. Phys.} {\bfseries 83}, 793 (2011)},
[{{\ttfamily  arXiv:1102.4014}}].


\bibitem{Cardoso:2016rao}
V.~Cardoso, E.~Franzin, and P.~Pani,
\emph{{Is the gravitational-wave ringdown a probe of the event horizon?}}
{\emph{ Phys. Rev. Lett.} {\bfseries 116}, 171101  (2016)},
[erratum: \emph{ Phys. Rev. Lett.}  {\bfseries 117} , 089902 (2016)]
[{{\ttfamily  arXiv:1602.07309}}].



\bibitem{Jusufi:2020odz}
K.~Jusufi, M.~Azreg-A\"\i{}nou, M.~Jamil, S.-W.~Wei, Q.~Wu, and A.-Z.~Wang,
\emph{{Quasinormal modes, quasiperiodic oscillations, and the shadow of rotating regular black holes in nonminimally coupled Einstein-Yang-Mills theory}},
{\emph{ Phys. Rev. D}  {\bfseries 103},  024013 (2021)},
[{{\ttfamily  arXiv:2008.08450}}].


\bibitem{Cheung:2021bol}
M.~H.~Y.~Cheung, K.~Destounis, R.~P.~Macedo, E.~Berti, and V.~Cardoso,
\emph{{Destabilizing the Fundamental Mode of Black Holes: The Elephant and the Flea}},
{\emph{ Phys. Rev. Lett.}  {\bfseries 128}, 111103  (2022)},
[{{\ttfamily arXiv:2111.05415}}].

\bibitem{Kristensen:2015qq}
P.~T.~Kristensen, R.-C.~Ge, and S.~Hughes,
\emph{{Normalization of quasinormal modes in leaky optical cavities and plasmonic resonators}},
{\emph{ Physical Review A},
	{\bfseries 92},
	053810 (2015)},
[{{\ttfamily  arXiv:1501.05938}}].

\bibitem{Seahra:2005wk}
S.~S.~Seahra,
\emph{Ringing the Randall-Sundrum braneworld: Metastable gravity wave bound states},
{\emph{Phys. Rev. D} {\bfseries  72}, 066002 (2005)},
[{{\ttfamily arXiv:hep-th/0501175}}].


\bibitem{Seahra:2005iq}
S.~S.~Seahra,
\emph{Metastable massive gravitons from an infinite extra dimension},
{\emph{Int. J. Mod. Phys. D} {\bfseries  14},  2279 (2005)},
[{{\ttfamily arXiv:hep-th/0505196}}].




\bibitem{Akarsu:2020yqa}
\"O.~Akarsu, N.~Kat\i{}rc\i{}, S.~Kumar, R.~C.~Nunes, B.~\"Ozt\"urk and S.~Sharma,
\emph{Rastall gravity extension of the standard $\Lambda $CDM model: theoretical features and observational constraints},
{\emph{Eur. Phys. J. C} {\bfseries 80} 1050 (2020)},
[{{\ttfamily arXiv:2004.04074}}].




\bibitem{Ciftci:2003As}
H. Ciftci, R. L. Hall, and N. Saad,
\emph{Asymptotic iteration method for eigenvalue problems},
{\emph{Journal of Physics A},  {\bfseries 36},  11807 (2003)},
[{{\ttfamily arXiv:math-ph/0309066}}].


\bibitem{ciftci:2005co}
H. Ciftci, R. L. Hall, and N. Saad,
\emph{Construction of exact solutions to eigenvalue problems by the asymptotic iteration method},
{\emph{Journal of Physics A: Mathematical and General},
 {\bfseries 38}, 1147 (2005)},
[{{\ttfamily arXiv:math-ph/0412030}}].



\bibitem{Cho:2011sf}
H.-T.~Cho, A.~S.~Cornell, J.~Doukas, T.-R.~Huang, and W.~Naylor,
\emph{A New Approach to Black Hole Quasinormal Modes: A Review of the Asymptotic Iteration Method},
{\emph{Adv. Math. Phys.} {\bfseries  2012}, 281705 (2012)},
[{{\ttfamily arXiv:1111.5024}}].



\bibitem{Csaki:2000pp}
C.~Csaki, J.~Erlich, and T.~J. Hollowood, \emph{{Quasilocalization of gravity by
		resonant modes}},
{\emph{Phys. Rev. Lett.}
	{\bfseries 84}, 5932 (2000)},
[{{\ttfamily arXiv:hep-th/0002161}}].

\end{thebibliography}
\end{document}